\documentclass[aps,prl,superscriptaddress,twocolumn,longbibliography]{revtex4-1}
\setcitestyle{super}
\usepackage{amsfonts,amssymb,amscd,amsthm}
\usepackage{graphicx}
\usepackage{mathrsfs}
\usepackage[intlimits]{amsmath}
\usepackage[colorlinks, citecolor=red]{hyperref}
\usepackage{etoolbox}
\usepackage{lineno}
\begin{document}

\title{Observation of Supersymmetry and its Spontaneous Breaking in a Trapped Ion Quantum Simulator}

\author{M.-L. Cai}
\thanks{These authors contribute equally to this work}
\affiliation{Center for Quantum Information, Institute for Interdisciplinary Information Sciences, Tsinghua University, Beijing, 100084, PR China}
\affiliation{HYQ Co., Ltd., Beijing, 100176, PR China}

\author{Y.-K. Wu}
\thanks{These authors contribute equally to this work}
\affiliation{Center for Quantum Information, Institute for Interdisciplinary Information Sciences, Tsinghua University, Beijing, 100084, PR China}

\author{Q.-X. Mei}
\affiliation{Center for Quantum Information, Institute for Interdisciplinary Information Sciences, Tsinghua University, Beijing, 100084, PR China}

\author{W.-D. Zhao}
\affiliation{Center for Quantum Information, Institute for Interdisciplinary Information Sciences, Tsinghua University, Beijing, 100084, PR China}

\author{Y. Jiang}
\affiliation{Center for Quantum Information, Institute for Interdisciplinary Information Sciences, Tsinghua University, Beijing, 100084, PR China}

\author{L. Yao}
\affiliation{Center for Quantum Information, Institute for Interdisciplinary Information Sciences, Tsinghua University, Beijing, 100084, PR China}
\affiliation{HYQ Co., Ltd., Beijing, 100176, PR China}

\author{L. He}
\affiliation{Center for Quantum Information, Institute for Interdisciplinary Information Sciences, Tsinghua University, Beijing, 100084, PR China}

\author{Z.-C. Zhou}
\affiliation{Center for Quantum Information, Institute for Interdisciplinary Information Sciences, Tsinghua University, Beijing, 100084, PR China}

\affiliation{Beijing Academy of Quantum Information Sciences, Beijing 100193, PR China}

\author{L.-M. Duan}
\email{lmduan@mail.tsinghua.edu.cn}
\affiliation{Center for Quantum Information, Institute for Interdisciplinary Information Sciences, Tsinghua University, Beijing, 100084, PR China}

\begin{abstract}
Supersymmetry (SUSY) helps solve the hierarchy problem in high-energy physics and provides a natural groundwork for unifying gravity with other fundamental interactions. While being one of the most promising frameworks for theories beyond the Standard Model, its direct experimental evidence in nature still remains to be discovered. Here we report experimental realization of a supersymmetric quantum mechanics (SUSY QM) model, a reduction of the SUSY quantum field theory for studying its fundamental properties, using a trapped ion quantum simulator. We demonstrate the energy degeneracy caused by SUSY in this model and the spontaneous SUSY breaking. By a partial quantum state tomography of the spin-phonon coupled system, we explicitly measure the supercharge of the degenerate ground states, which are superpositions of the bosonic and the fermionic states. Our work demonstrates the trapped-ion quantum simulator as an economic yet powerful platform to study versatile physics in a single well-controlled system.
\end{abstract}

\maketitle

\section{Introduction}
Supersymmetry (SUSY) is a quantum mechanical symmetry that unifies the space-time and the internal degree of freedom of elementary particles \cite{weinberg2000quantum,aitchison2007supersymmetry}. In an SUSY theory, bosons and fermions have one-to-one correspondence with the same mass, which leads to a cancellation in the Lagrangian of the Higgs particle and thus helps solve the hierarchy problem or the ultraviolet divergence in Standard Model \cite{aitchison2007supersymmetry}. However, in reality, this cancellation in Higgs mass is not exact and no SUSY partners of known particles have been discovered at the current energy scales, which requires the spontaneous breaking of SUSY \cite{WITTEN1981513,aitchison2007supersymmetry}. Since it is mathematically daunting to decide if SUSY is spontaneously broken in a quantum field theory, supersymmetric quantum mechanics (SUSY QM) \cite{SUSYQM2004,wasay2016supersymmetric} has been proposed as a toy model for understanding this crucial property \cite{WITTEN1981513}. In an SUSY QM model, Hamiltonians have nonnegative eigenvalues $E\ge 0$ with all the $E>0$ levels being degenerate, which is a consequence of the boson-fermion correspondence \cite{WITTEN1982253}. On the other hand, there can also be $E=0$ bosonic or fermionic ground states which are vacuum states annihilated by supercharges, the generators of SUSY, and do not necessarily have the boson-fermion correspondence. This can be characterized by the Witten index, the number of $E=0$ bosonic states minus that of the fermionic states \cite{WITTEN1982253}. If such $E=0$ levels do not exist, the SUSY is said to be spontaneously broken as the degenerate ground states with positive energies transform between bosons and fermions by the supercharge rather than being invariant \cite{WITTEN1982253}.

Although the SUSY theory has inspired wide applications in fields like optics \cite{PhysRevLett.110.233902}, condensed matter physics \cite{efetov1999supersymmetry}, quantum chaos \cite{efetov1999supersymmetry} or even outside physics \cite{bardoscia2021physics}, whether it correctly describes the physical world has not yet been determined by the state-of-the-art high energy physics experiments \cite{aaboud2018search,sirunyan2019search} or other indirect experimental evidences \cite{PhysRevLett.126.141801,planck2016}. However, many theoretical proposals already exist \cite{Hirokawa2015,tomka2015supersymmetry,PhysRevB.92.245444,minavr2020kink,gharibyan2021toward} to examine its effects by quantum simulation \cite{georgescu2014quantum,cirac2012goals}. As one of the leading platforms for quantum information processing with long coherence time, convenient initialization and readout, as well as accurate laser or microwave control \cite{RevModPhys.75.281,PhysRevLett.113.220501,PhysRevLett.117.060504,PhysRevLett.117.060505,wang2021single}, ion trap has demonstrated the quantum simulation of various phenomena such as quantum phase transitions \cite{monroe2021programmable,Rabi_model}, many-body dynamics \cite{monroe2021programmable}, relativistic effects \cite{Gerritsma2010} and quantum field theories \cite{kokail2019self}. In this work, we report experimental realization of a prototypical SUSY QM model \cite{Hirokawa2015} in a trapped ion quantum simulator and demonstrate the spontaneous SUSY breaking in this model. The SUSY QM model is realized by tuning suitable parameters in a quantum Rabi model (QRM) Hamiltonian \cite{Rabi_model,PhysRevX.8.021027}. Through a combination of the state-of-the-art manipulation techniques for spin and phonon states in ion trap, and the joint spin-phonon state tomography scheme we develop that has not been achieved before, we explicitly demonstrate the characteristic signatures of the SUSY theory by measuring the energy degeneracy between the bosonic and the fermionic states, and the nonvanishing supercharges of the degenerate ground states in the spontaneous SUSY breaking case.

\section{Results}
\begin{figure}[htbp]
   \includegraphics[width=\linewidth]{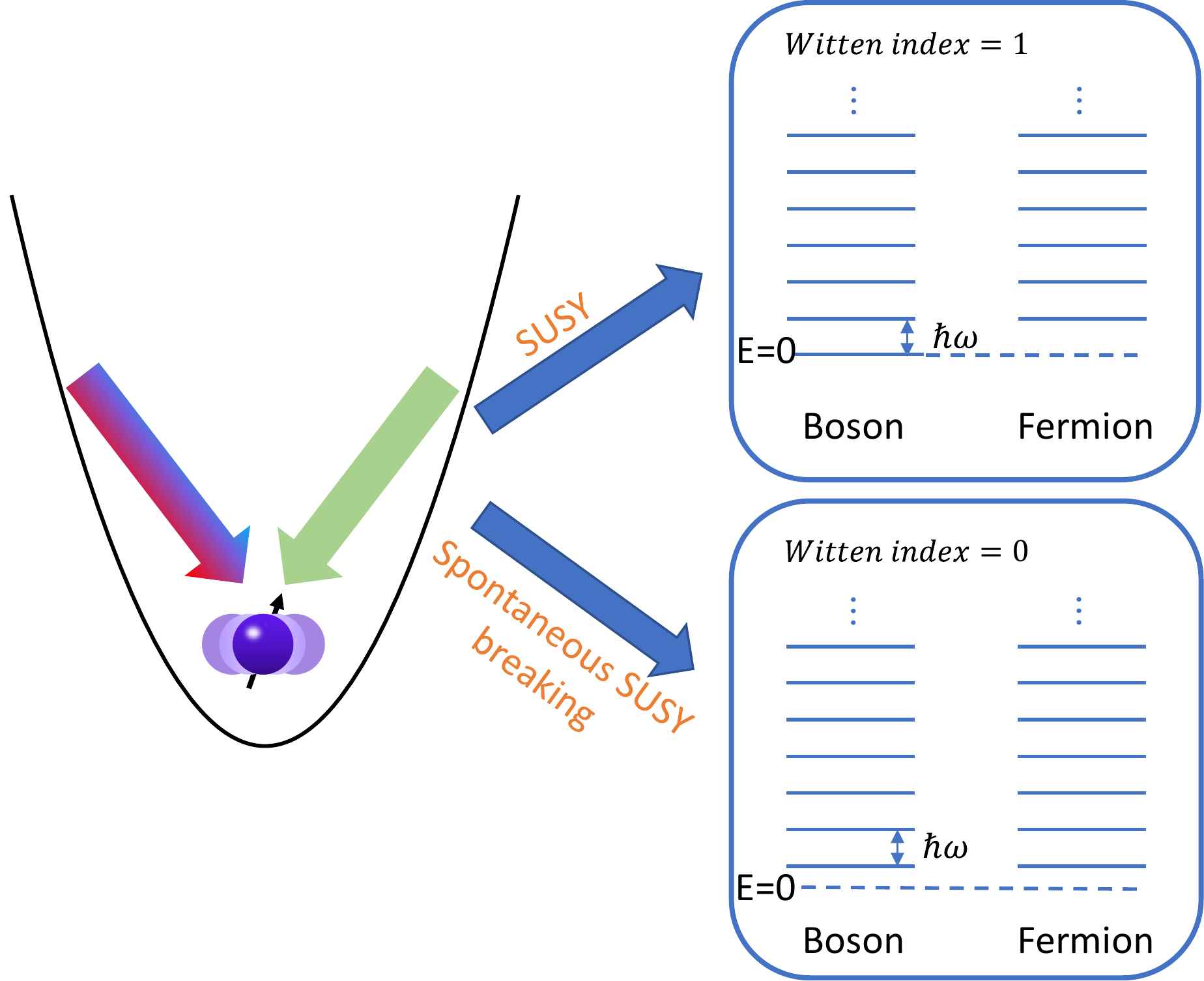}
   \caption{\textbf{Quantum Simulation of SUSY and its spontaneous breaking.} We simulate the QRM Hamiltonian by coupling the internal states of an ion with its spatial oscillation using bichromatic Raman laser beams. Under suitable parameters, the system possesses SUSY and there is a unique $E=0$ bosonic ground state with all the other excited states being double-degenerate for bosonic and fermionic states. This corresponds to a Witten index of one. When moving to a different set of parameters, the Hamiltonian is still supersymmetric, but this symmetry spontaneously breaks and the ground states are no longer unique but obtains non-vanishing supercharges and positive energies. The Witten index in this case is zero.}
   \label{fig:scheme}
\end{figure}
Our experimental scheme is sketched in Fig.~\ref{fig:scheme}. We create a QRM Hamiltonian with bichromatic Raman laser beams on a single trapped ${}^{171}\mathrm{Yb}^+$ ion \cite{Rabi_model,PhysRevX.8.021027} (see Methods)
\begin{equation}
H=\frac{\omega_s}{2} \sigma_z + \omega \left(a^\dag a+\frac{1}{2}\right) + g\sigma_x(a+a^\dag) + \frac{g^2}{\omega},
\label{eq:QRM}
\end{equation}
where $\sigma_z$, $\sigma_x$ denote the components of the Pauli spin operator, $a$ ($a^\dag$) the bosonic annihilation (creation) operator for the phonon mode, and the last term a renormalization constant \cite{Hirokawa2015}. This system possesses two supersymmetric points. One is at $g=0$ and $\omega_s=\omega$ where the system is as simple as a spin and an oscillator without coupling. This becomes evident if we write $\sigma_z=2f^{\dagger}f-1$ and $\sigma_x=f+f^{\dagger}$ where $f$ ($f^{\dagger}$) denotes the annihilation (creation) operator of a fermionic mode. The excited states $|\uparrow\rangle|n\rangle$ and $|\downarrow\rangle|n+1\rangle$ ($n\ge 0$) are degenerate fermionic and bosonic states ($|\uparrow\rangle$ and $|\downarrow\rangle$ have fermionic occupation number of $1$ and $0$, respectively) with positive energy $E_{n+1}=(n+1)\omega$, while the unique ground state $|\downarrow\rangle|0\rangle$ has zero vacuum energy $E_0=0$ owing to the cancellation between the spin (fermionic) energy $\omega_s$ and the phonon (bosonic) energy $\omega$.

\begin{figure*}[htbp]
   \includegraphics[width=\linewidth]{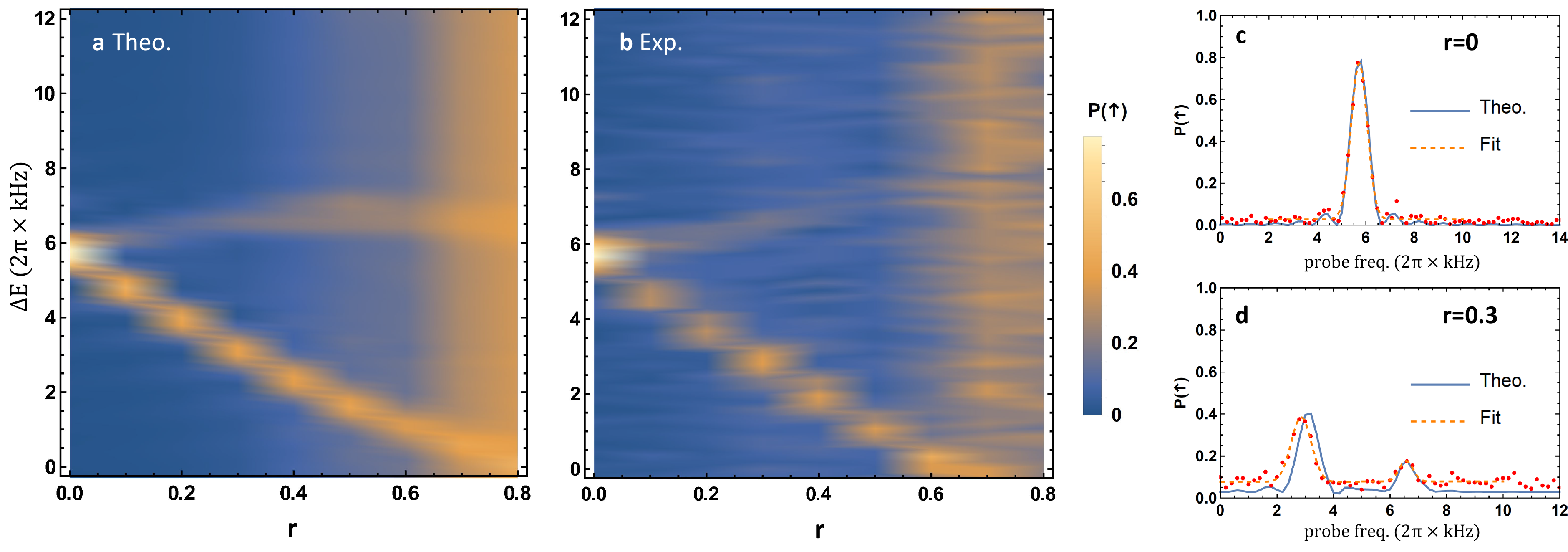}
   \caption{\textbf{QRM energy spectrum and SUSY.} \textbf{a} Theoretical energy gaps between the ground state and the first two excited states as a function of $r$ in Eq.~(\ref{eq:H_spectrum}). \textbf{b} Experimentally measured energy gaps. At $r=0$, the system is supersymmetric with a unique bosonic ground state and degenerate first and second excited states for the bosonic and the fermionic modes. As $r\to 1$, the Hamiltonian approaches the other supersymmetric point, but the unique ground state disappears and the lowest bosonic and fermionic states become degenerate, which indicates a spontaneous SUSY breaking. \textbf{c}, \textbf{d} The resonant signals probed by a weak pulse for energy gaps at $r=0$ and $r=0.3$, respectively. The solid blue curves are the theoretical results and the red dots are the measured data which are fitted by multiple Gaussian functions as the dashed orange curves to extract the peak locations.}
   \label{fig:susy_spectrum}
\end{figure*}

The other more interesting SUSY point appears if we turn on the spin-phonon coupling $g$ and set $\omega_s=0$ (see Supplementary Note 1). In this special situation, while the Hamiltonian is still supersymmetric, its ground state is not, which indicates a spontaneous SUSY breaking. We connect these two supersymmetric points by the path
\begin{equation}
H=\frac{(1-r)\omega}{2} \sigma_z + \omega \left(a^\dag a+\frac{1}{2}\right) + rg_m\sigma_x(a+a^\dag) + \frac{(rg_m)^2}{\omega},
\label{eq:H_spectrum}
\end{equation}
which corresponds to $\omega_s=(1-r)\omega$ and $g=r g_m$ in the QRM. The energy spectrum for this system under $\omega=g_m=2\pi\times 5.73\,$kHz is plotted in Fig.~\ref{fig:susy_spectrum}a, b with $r$ ranging from zero to 0.8. At $r=0$, the first and the second excited states are degenerate, with a finite energy gap of $\omega$ from the unique ground state. As $r$ increases, the degeneracy between the two excited states is lifted while the gap between the ground state and the first excited state shrinks. To probe this energy spectrum, we first prepare the ground state at the desired $r$ through an adiabatic passage; then we apply a weak probe pulse $H_p=\Omega_p \sigma_x \cos\omega_p t$ with the QRM Hamiltonian turned on and measure the change in the spin population \cite{PhysRevX.8.021027} (see Methods). By scanning $\omega_p$, we obtain the energy gap between the ground and the excited states as shown in Fig.~\ref{fig:susy_spectrum}c, d as two examples.

In the above process, the energy gap between the ground state and the first excited state closes as $r\to 1$, which seems to violate the adiabatic condition when preparing the ground state. Nevertheless, these two states locate in different symmetry branches of the QRM under parity transform $\sigma_z e^{i\pi a^\dag a}$, which commutes with the QRM Hamiltonian. Therefore the two states will not evolve into each other and the adiabatic condition can still hold. However, there is another problem that as $r$ increases, the coupling between these two lowest levels by $H_p$ becomes weaker, which leads to a reduction in the resonant signal even for the ideal state as illustrated in Fig.~\ref{fig:susy_spectrum}a at large $r$. Consequently, it is still difficult to accurately determine the energy degeneracy near $r=1$, the nontrivial spontaneous SUSY breaking point. For this purpose, we explicitly measure the energy expectation values of the two ground states at $r=1$.

To prepare these two states, we fix $\omega_s=0$ in Eq.~(\ref{eq:QRM}) and gradually turn up $g$ from zero to $g_m$ (see Methods). The two initial ground states $|\uparrow\rangle|0\rangle$ and $|\downarrow\rangle|0\rangle$ will then adiabatically evolve into the ground states at the spontaneous SUSY breaking point $|\psi_\pm \rangle=(|+\rangle|-g_m/\omega\rangle \pm |-\rangle|g_m/\omega\rangle)/\sqrt{2}$ which are ideally two Schr\"odinger's cat states.

\begin{figure*}[htbp]
   \includegraphics[width=\linewidth]{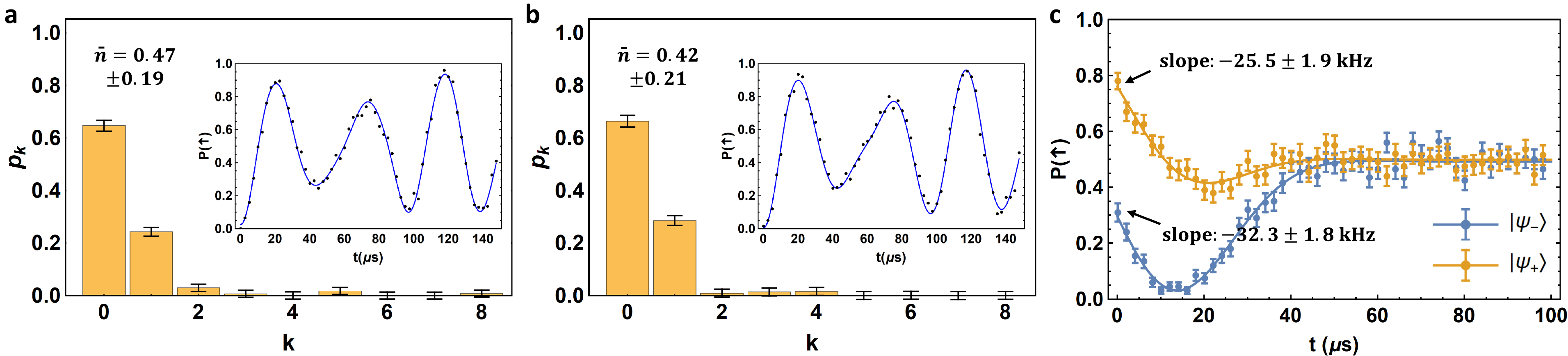}
   \caption{\textbf{Measuring ground state energies at spontaneous SUSY breaking point.} \textbf{a}, \textbf{b} Phonon number distribution of the ground states $|\psi_{+}\rangle$ and $|\psi_{-}\rangle$, respectively, by fitting the time evolution under a blue-sideband driving as shown in the inset. The estimated average phonon numbers are $\bar{n}_+=0.47\pm0.19$ and $\bar{n}_-=0.42\pm0.21$, respectively. \textbf{c} Measuring $\langle\sigma_x(a+a^\dag)\rangle$ by fitting the spin-up state population under a spin-dependent force. The dynamics can be fitted by an analytic formula (solid curves, see Methods) and then the slope at $t=0$ can be extracted as $(-25.5\pm1.9)\,$kHz and $(-32.3\pm1.8)\,$kHz for $|\psi_{+}\rangle$ and $|\psi_{-}\rangle$, respectively. Each data point is averaged over $N=200$ experimental shots. Error bars represent one standard deviation. Combining the two results, we measure the average energy of the two ground states as $E_+=2\pi\times(7.3\pm 1.9)\,$kHz and $E_-=2\pi\times(5.2\pm 2.1)\,$kHz with the ideal value $E_0=\omega/2=2\pi\times 5\,$kHz. The two levels are degenerate within about one standard deviation.}
   \label{fig:ground_energy}
\end{figure*}
Now to evaluate the expectation value of the Hamiltonian $H=\omega(a^\dag a + 1/2)+g_m\sigma_x(a+a^\dag)+g_m^2/\omega$, we only need to measure the average phonon number $\langle a^\dag a\rangle$ and a spin-phonon coupling term $\langle\sigma_x(a+a^\dag)\rangle$. The former can be derived by the standard procedure of first resetting the spin state to $|\downarrow\rangle$ and then driving the phonon blue-sideband to fit the phonon number population from the spin dynamics \cite{RevModPhys.75.281}, as shown in Fig.~\ref{fig:ground_energy}a, b. As for the second term, we apply a spin-dependent force $H_{\mathrm{SDF}}=(-\Omega_p/2)\sigma_y(a+a^\dag)$ and measure the evolution of $\sigma_z(t)$ by observing that $e^{iH_{\mathrm{SDF}}t} \sigma_z e^{-iH_{\mathrm{SDF}}t}$ has a linear term in $t$ as $\Omega_p t \sigma_x(a+a^\dag)$ \cite{Gerritsma2010}. Therefore, after preparing $|\psi_{\pm}\rangle$, we adjust the parameters of the bichromatic Raman laser beams to create this spin-dependent force and fit $\langle\sigma_z(t)\rangle$ to extract the slope at $t=0$ (see Methods), from which we obtain $\langle\sigma_x(a+a^\dag)\rangle$, as shown in Fig.~\ref{fig:ground_energy}c. Combining these two results, we obtain $E_+=2\pi\times(7.3\pm 1.9)\,$kHz and $E_-=2\pi\times(5.2\pm 2.1)\,$kHz for $|\psi_+\rangle$ and $|\psi_-\rangle$ respectively at the parameters $\omega=2\pi\times 10\,$kHz and $g_m=2\pi\times 5.73\,$kHz. The two levels are degenerate within about one standard deviation, and also agree well with the theoretical value of $E_0=\omega/2=2\pi\times 5\,$kHz. Note that the error bar here, mainly caused by fitting the phonon distribution, is much smaller than the gap $\omega$ to higher levels. Such a double-degeneracy of the ground state with positive energy clearly indicates the spontaneous SUSY breaking.

\begin{figure*}[htbp]
   \includegraphics[width=\linewidth]{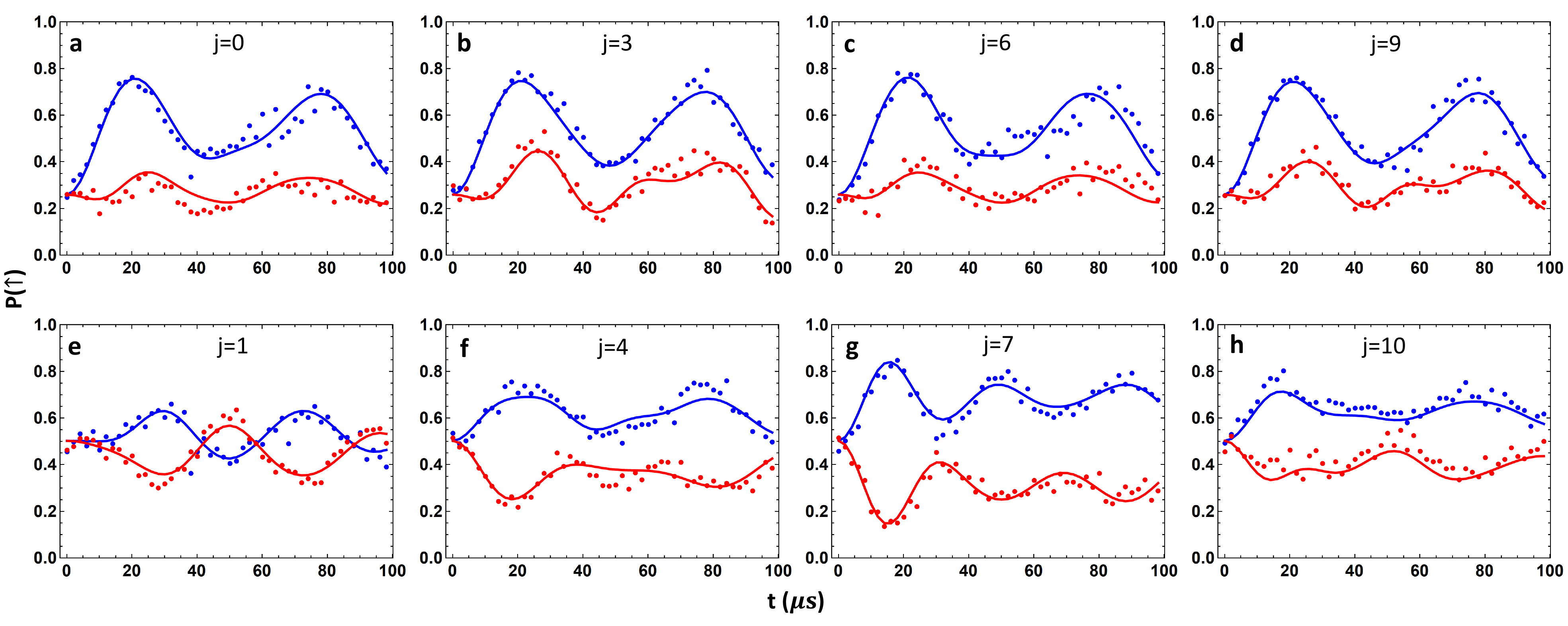}
   \caption{\textbf{Measuring ground state supercharges.} We reconstruct the phonon density matrix with the spin projected to $\sigma_z=\pm 1$ or $\sigma_y=\pm 1$ by maximum likelihood estimation. \textbf{a}-\textbf{d} The measured spin dynamics under blue-sideband (blue dots) and red-sideband (red dots) driving and those predicted by the best-fitted density matrix (blue and red curves) for $|\psi_-\rangle$ after four typical phonon displacements with $\sigma_z$ projected to $\pm 1$. Each data point is the average of 400 experimental shots, half of which have a $\sigma_z$ gate before the measurement sequence to remove the off-diagonal spin terms. \textbf{e}-\textbf{h} Similar plots for $|\psi_-\rangle$ after four typical phonon displacements with $\sigma_y$ projected to $\pm 1$. Here we choose $\omega=2\pi\times 10\,$kHz and $g_m=2\pi\times 5.43\,$kHz. The displacement operators $D(\beta_j)$ are characterized by $\beta_j=i\beta e^{2\pi i j / N}$ where $\beta=0.687$, $N=12$ and $j=0,\,1,\,\cdots,\, N-1$. The complete data for all the displacements and both $|\psi_-\rangle$ and $|\psi_+\rangle$ are presented in Supplementary Note 2. From the fitted density matrices, we get $\langle\sigma_z\otimes A\rangle = 0.201 \pm 0.013$ and $\langle\sigma_y\otimes B\rangle = 0.300 \pm 0.009$ for $|\psi_-\rangle$ and $\langle\sigma_z\otimes A\rangle = -0.163\pm 0.017$ and $\langle\sigma_y\otimes B\rangle = -0.349 \pm 0.011$ for $|\psi_+\rangle$ where error bars are estimated by Monte Carlo sampling (see Methods).}
   \label{fig:supercharge}
\end{figure*}
In quantum field theory, $E_0>0$ has the nontrivial meaning of a positive vacuum energy. However, in quantum mechanics, a constant in the Hamiltonian only contributes to a global phase and can always be discarded. Therefore we would prefer more direct evidences of the spontaneous SUSY breaking by measuring the supercharge, which is the generator of the SUSY that transforms between bosons and fermions. An SUSY-invariant vacuum will be annihilated by the supercharge, while in the case of spontaneous SUSY breaking, the degenerate ground states can take nonzero supercharges. For the Hamiltonian of Eq.~(\ref{eq:H_spectrum}) with $r=1$, the supercharge can be defined as \cite{Hirokawa2015} (see Supplementary Note 1 for details)
\begin{equation}
\begin{aligned}
Q=&-\frac{\sigma_{z}}{2}\left[V_{+} \sqrt{a^{\dagger} a+\frac{1}{2}} V_{+}+V_{-} \sqrt{a^{\dagger} a +\frac{1}{2}} V_{-}\right]  \\
&-i\frac{\sigma_{y}}{2}\left[V_{+} \sqrt{a^{\dagger} a +\frac{1}{2}} V_{+}-V_{-} \sqrt{a^{\dagger} a +\frac{1}{2}} V_{-}\right] \\
\equiv&\sigma_z\otimes A+\sigma_y\otimes B,
\end{aligned}
\end{equation}
where $V_{\pm}\equiv \mathrm{exp}[\pm (g_m/\omega) (a^\dag - a)]$ are displacement operators, and we define two phonon operators $A$ and $B$ to simplify the expression. We have taken out a coefficient $\sqrt{\omega}$ to make the supercharge dimensionless. Ideally, $|\psi_\pm\rangle$ are eigenstates of $Q$ with eigenvalues of $\mp 1/\sqrt{2}$.

To evaluate $\langle \sigma_z \otimes A\rangle$, in principle we can decompose the spin-phonon density matrix as $\rho=|\uparrow\rangle\langle \uparrow| \otimes \rho_{\uparrow\uparrow} + |\uparrow\rangle\langle \downarrow| \otimes \rho_{\uparrow\downarrow} + |\downarrow\rangle\langle \uparrow| \otimes \rho_{\downarrow\uparrow} + |\downarrow\rangle\langle \downarrow| \otimes \rho_{\downarrow\downarrow}$ where $\rho_{\alpha}$ ($\alpha=\uparrow\uparrow,\,\uparrow\downarrow,\,\downarrow\uparrow,\,\downarrow\downarrow$) are operators on the phonon states. Then we have $\langle \sigma_z \otimes A\rangle = \mathrm{tr}[(\rho_{\uparrow\uparrow} - \rho_{\downarrow\downarrow}) A]$. Naively, to obtain $\rho_{\uparrow\uparrow}$ and $\rho_{\downarrow\downarrow}$ separately, we can first measure $\sigma_z$ and then reset the spin state as an ancilla for quantum state tomography on the phonon state \cite{RevModPhys.75.281}, conditioned on the measurement outcome $\sigma_z=\pm 1$. However, in ion trap experiments, this measurement of the spin state typically takes hundreds of microseconds to accumulate sufficient photon counts, during which the decoherence of the phonon state may not be ignored. This suggests that we shall incorporate the spin states into the tomography of the phonon states. Rather than resetting the spin state to $|\downarrow\rangle$, we remove its off-diagonal term by applying random $\sigma_z$ operations but keep the population in the $|\uparrow\rangle$ and $|\downarrow\rangle$ basis (see Methods). Then we follow the standard procedure of phonon quantum state tomography by applying displacements in different directions and blue/red sideband driving \cite{RevModPhys.75.281,PhysRevLett.116.140402}. Through maximum likelihood estimation (MLE) \cite{PhysRevA.63.040303}, we can reconstruct the joint spin-phonon density matrix in the diagonal spin basis, that is, $\rho_{\uparrow\uparrow}$ and $\rho_{\downarrow\downarrow}$. Some typical fitting results for various displacement directions and blue/red sidebands are presented in Fig.~\ref{fig:supercharge}a-d. Similarly, by first rotating the $\sigma_y$ basis to the $\sigma_z$ basis, we can reconstruct the partial information required to evaluate $\langle\sigma_y\otimes B\rangle$, as shown in Fig.~\ref{fig:supercharge}e-h. Combining the two results, we get $Q_-=0.501\pm0.016$ and $Q_+=-0.512\pm 0.020$ after correcting a phase in the phonon state due to the slow drift of the trap frequency (see Methods and Supplementary Note 2). The measured supercharges are significantly away from zero, which suggests that the ground states are superposition of bosonic and fermionic states rather than an SUSY-invariant vacuum. Note that here the error bars only account for the statistical errors of MLE in finding the best-fitted density matrices for the prepared states (see Methods), while imperfect state preparation and measurement can still result in a systematic deviation from the ideal values of $\pm 0.707$. For example, a slow fluctuation in the trap frequency of $2\pi\times 0.5\,$kHz during the joint tomography, which takes tens of minutes, can already reduce the supercharges to about $\pm 0.5$ (see Methods). Also note that, while here we are only interested in the $\sigma_z$ and $\sigma_y$ parts of the spin for measuring supercharges, our method can be extended to include the $\sigma_x$ part as well. Combining these results altogether one will be able to perform full quantum state tomography of the joint spin-phonon system.

To sum up, we have demonstrated the quantum simulation of an SUSY QM model using a single trapped ion. By measuring the energy spectrum of the QRM along a carefully chosen path, we illustrate the characteristic properties in the degeneracy of the energy levels, whether the SUSY spontaneously breaks or not. We adiabatically prepare the degenerate ground states at the spontaneous SUSY breaking point and explicitly measure the expectation value of the energy and the supercharge, from which the spontaneous SUSY breaking can be verified experimentally. Recently, quantum simulation using up to 16 ions and 16 phonon modes has been demonstrated \cite{mei2021experimental}, which should allow the demonstration of some lattice quantum field theory models using spin-boson coupled systems, similar to Ref.~\cite{kokail2019self} for spin models. To follow this direction to further explore the SUSY theory, suitable SUSY lattice models need to be designed to fit into the ion trap setup, and suitable observables should be chosen to characterize its properties: while the joint quantum state tomography scheme we develop here can directly be generalized to the multi-ion case to reconstruct the joint state of any particular collective phonon mode with any particular spin, and provide some useful information, in general the full tomography of all the spins and all the phonon modes will be exponentially more challenging. Nevertheless, as we show in this experiment, partial information about the system may suffice to demonstrate the spontaneous SUSY breaking or other properties of concern. Our work showcases the suitability of ion trap as an economic but powerful platform for simulating diverse physics through its unparalleled controllability.

\section{Methods}
\subsection{Experimental setup}
\label{app:setup}
Our experimental setup is a single ${}^{171}\mathrm{Yb}^+$ ion in a linear Paul trap. The spin state is encoded in the $|\downarrow \rangle=|{}^{2}S_{1/2},F=0,m_F=0\rangle$ and the $|\uparrow \rangle=|{}^{2}S_{1/2},F=1,m_F=0\rangle$ levels of the ion with atomic transition frequency around $\omega_q=2\pi\times 12.643\,$GHz, and we exploit a radial oscillation mode with secular frequency around $\omega_x=2\pi\times 2.35\,$MHz.

We use counter-propagating $355\,$nm pulsed laser beams with a specially designed repetition rate around $2\pi \times 118.415\,$MHz and a bandwidth of about $200\,$GHz to manipulate the ion through Raman transition. Two acousto-optic modulators (AOMs) are used to fine-tune the frequency, the phase and the amplitude of the laser beams.

Before each experiment, we initialize the phonon state to $|0\rangle$ through sideband cooling using $355\,$nm laser. Then we initialize the spin state by optical pumping of $369\,$nm laser with a $935\,$nm laser to repump the population leaked to the ${}^{2}D_{3/2}$ levels. More details about our setup can be found in our previous work \cite{Rabi_model}.

\subsection{Simulating quantum Rabi model}
\label{app:QRM}
We follow the scheme of Refs.~\cite{Rabi_model,PhysRevX.8.021027} to simulate the QRM for an ion with qubit frequency $\omega_q$ in a harmonic trap with trap frequency $\omega_x$. By applying bichromatic Raman laser beams along the $x$ direction, we get red (blue) sideband Hamiltonian $H_{r(b)}=\Omega_{r(b)} \sigma_x \cos(k_{r(b)} x - \omega_{r(b)} t + \phi_{r(b)})$ with small detuning $\delta_r=\omega_q-\omega_x-\omega_r$ ( $\delta_b=\omega_q+\omega_x-\omega_b$) to the red (blue) sideband, respectively. By setting $\delta_r=\omega_s-\omega$, $\delta_b=\omega_s+\omega$ and $\Omega_r=\Omega_b=2g/\eta$ where $\eta=k\sqrt{\hbar/2m\omega_x}$ is the Lamb-Dicke parameter, we get the QRM Hamiltonian of Eq.~(\ref{eq:QRM}) in the interaction picture of $H_0=(\omega_q-\omega_s)\sigma_z/2+(\omega_x-\omega)a^\dag a$.

\subsection{Measuring excitation spectrum}
\label{app:spectrum}
After preparing the ground state of the system (see below), we apply a weak probe $H_p=\Omega_p \sigma_x \cos\omega_p t$ with $\Omega_p=2\pi\times 0.17\,$kHz (much smaller than the typical energy scale $\omega$ and $g_m$ of the system) and a duration $\tau=1000\,\mu$s to measure the excitation spectrum of the system \cite{PhysRevX.8.021027}, whose peaks give the energy gaps to the excited states. Note that the QRM Hamiltonian is realized in an interaction picture as mentioned above, in which the effective probe frequency is $\omega_p-\omega_q+\omega_s$. This is what we plot as the horizontal axes in Fig.~\ref{fig:susy_spectrum}c, d.

\subsection{Adiabatic evolution}
\label{app:adiabatic}
In principle we can use arbitrary adiabatic passages to prepare the ground states so long as the adiabatic condition is satisfied. However, the change in the AC Stark shift must be taken into account when we tune the laser frequency or intensity, which may accumulate into a non-negligible phase during the slow quench.

To prepare the ground state of Eq.~(\ref{eq:H_spectrum}), we first set $\omega_s=(1-r)\omega$ and $g=r g_m$ at the desired values, and choose the phonon frequency to be $10\omega$ which is significantly larger than the coupling $g$. We initialize the system to be $|\downarrow\rangle|0\rangle$ which is close to the ground state in this case. Here if we apply a weak probe to measure the excitation spectrum as mentioned above, since the spin-phonon coupling can be neglected, the resonant signal is expected to locate at $\omega_p=\omega_q$, that is, the carrier frequency. This allows us to calibrate the AC Stark shift under the coupling strength we use. Now we can adiabatically turn down the phonon frequency following an exponential function \cite{PhysRevX.8.021027} $(9 e^{-t/T}+1)\omega$ with $T=40\,\mu$s and an overall quench time of $400\,\mu$s. From numerical simulation, the non-adiabatic excitation can be below 1\% for $r$ ranging from 0 to 0.9.

As for the degenerate ground states used for measuring the average energy and the supercharge, we observe that in the case of $r=1$, the driving on the blue and the red sidebands are balanced. Our $355\,$nm pulsed laser has a specially designed repetition rate such that in this situation the AC Stark shift from these two sidebands largely cancel each other \cite{mei2021experimental}. For coupling as large as $2\pi\times 10\,$kHz, the measured AC Stark shift is still below $100\,$Hz. Therefore here we simply perform a linear quench in the coupling $g$ as $g(t)=g_m t/\tau$ with $\tau=200\,\mu$s. From numerical simulation we see that the error due to the violation of the adiabatic condition is far below 0.1\% and is negligible in our experiment.

\subsection{Measuring expectation value of Hamiltonian}
\label{app:expectation}
The Hamiltonian at the spontaneous SUSY breaking point $r=1$ contains a phonon term $\omega a^\dag a$ and a spin-phonon coupling term $g_m \sigma_x (a+a^\dag)$. The phonon term can be measured by fitting the phonon number distribution through the spin dynamics under a blue-sideband driving \cite{RevModPhys.75.281}. Details can be found in our previous work \cite{Rabi_model}.

To measure the spin-phonon coupling term, we observe that $\sigma_z(t)=U^\dag(t) \sigma_z U(t)=\cos[\Omega_p t(a+a^\dag)]\sigma_z + \sin[\Omega_p t(a+a^\dag)]\sigma_x$ has a linear term in $t$ as $\Omega_p t \sigma_x(a+a^\dag)$, where $U(t)=e^{i(\Omega_p t / 2)\sigma_y(a+a^\dag)}$ is the evolution under a spin-dependent force. Therefore we measure $\langle\sigma_z(t)\rangle$ under the spin-dependent force and fit its slope at $t=0$ to get $\langle\sigma_x(a+a^\dag)\rangle$. This can in principle be obtained by measuring only the initial part of the evolution, which, however, would require high measurement accuracy. Here we observe that for the ideal ground states $|\psi_\pm \rangle=(|+\rangle|-\beta\rangle \pm |-\rangle|\beta\rangle)/\sqrt{2}$ ($\beta=g_m/\omega$), we have $\langle\sigma_z(t)\rangle = e^{-\Omega_p^2 t^2 / 2} (\pm e^{-2\beta^2}-\sin 2\beta\Omega_p t)$ using the fact that displacement operator $D(\gamma)=e^{-|\gamma|^2/2} e^{\gamma a^\dag} e^{-\gamma^* a}$ and thus $\langle\alpha|D(\gamma)|\beta\rangle=\mathrm{exp}[-(|\alpha|^2+|\beta|^2+|\gamma|^2)/2 +\gamma\alpha^*-\gamma^*\beta+\alpha^*\beta]$. Considering experimental imperfections, we fit the measured spin-up state population as $P_\uparrow(t)=A+B e^{-\Omega_p^2 t^2 / 2} (\pm e^{-2\beta^2}-\sin 2\beta\Omega_p t)$ with two fitting parameters $A$ and $B$, as shown in Fig.~\ref{fig:ground_energy}c. After fitting these two parameters, we can compute the slope at $t=0$ and evaluate the spin-phonon coupling term using $\Omega_p=2\pi\times 8.1\,$kHz.

\subsection{Partial spin-phonon state tomography}
\label{app:tomography}
Rather than first projecting the spin state to $\sigma_z=\pm 1$ and then perform the quantum state tomography for the phonon state conditionally, here we reconstruct the the desired $\rho_{\uparrow\uparrow}$ and $\rho_{\downarrow\downarrow}$ altogether. Different from the standard procedure for phonon state tomography where the spin state needs to be discarded \cite{RevModPhys.75.281}, here we only remove the off-diagonal terms of the spin state by inserting a $\sigma_z$ gate on half of the measurements. Then we follow the steps of the phonon state tomography by applying phonon displacements in various directions and measure the evolution of spin population under a driving on the blue or red sidebands \cite{RevModPhys.75.281}. Theoretically, after removing the spin coherence between $|\uparrow\rangle$ and $|\downarrow\rangle$, the spin-up state population for any quantum state under the blue or red sideband driving can be given by
\begin{equation}
\begin{aligned}
P_{b,\uparrow}(t)=&\frac{1}{2}\left[1+\sum_{n=0}^{N_{\mathrm{cut}}}(P^{\uparrow\uparrow}_{n+1} - P^{\downarrow\downarrow}_{n}) \cos \left(\Omega_{n, n+1}t\right) e^{-\gamma_{n} t}\right] \\
&+\frac{1}{2}P^{\uparrow\uparrow}_{0},
\end{aligned}
\end{equation}
and
\begin{equation}
\begin{aligned}
P_{r,\uparrow}(t)=& \frac{1}{2}\left[1+\sum_{n=0}^{N_{\mathrm{cut}}}(P^{\uparrow\uparrow}_{n} - P^{\downarrow\downarrow}_{n+1}) \cos \left(\Omega_{n, n+1}t\right) e^{-\gamma_{n} t}\right] \\ &-\frac{1}{2}P^{\downarrow\downarrow}_{0}.
\end{aligned}
\end{equation}

In principle, by fitting these curves, one can get complete information about $P^{\uparrow\uparrow}_{n}$ and $P^{\downarrow\downarrow}_{n}$, the diagonal elements in the spin-phonon joint density matrix. Then by applying different phonon displacements, the desired density matrices $\rho_{\uparrow\uparrow}$ and $\rho_{\downarrow\downarrow}$ can be obtained \cite{RevModPhys.75.281}.

In practice, this procedure of first extracting the coefficients $P^{\uparrow\uparrow}_{n}$ and $P^{\downarrow\downarrow}_{n}$ and then solving the original density matrix can be sensitive to noise in the measurement, and the obtained matrix may not satisfy the physical constraints such as being positive semidefinite with a trace of one. Therefore here we use maximum likelihood estimation \cite{PhysRevA.63.040303} to find the most probable density matrix that can generate all the measured blue-/red-sideband evolutions. We use the Hamiltonian $H_b = (\Omega_b / 2) \sigma_+ (a^\dag + \beta) + \sigma_- (a + \beta^*)$ and $H_r = (\Omega_r / 2) \sigma_- (a^\dag + \beta) + \sigma_+ (a + \beta^*)$ to drive the blue or red sidebands along with a phonon displacement $\beta$ at the same time \cite{PhysRevLett.116.140402}. We use $N=12$ displacements $D(\beta_j)$ with $\beta_j=i\beta e^{2\pi i j / N}$, $\beta=0.687$ and $j=0,\,1,\,\cdots,\, N-1$, and we set a phonon cutoff of $n_{\mathrm{cut}}=7$ for the reconstructed density matrices with only $10^{-9}$ population outside the truncated Hilbert space for the ideal states.

The projection to $\sigma_y=\pm 1$ can be obtained in a similar way by first rotating the $\sigma_y$ basis to the $\sigma_z$ basis. The complete data and fitting results are presented as Figs.~S1-S4 in Supplementary Note 2.

\subsection{Calibrate phonon state rotation error due to trap frequency drift}
\label{app:drift}
In Supplementary Figure 5 of Supplementary Note 2, we present the fidelity between the reconstructed density matrices and the ideal ones (with the off-diagonal spin terms discarded as described in the main text), versus a rotation angle $\theta$ applied on the phonon state $R(\theta)=e^{i\theta a^\dag a}$. As we can see, while $|\psi_-\rangle$ has the highest fidelity near $\theta=0$, the best result for $|\psi_+\rangle$ is achieved at a finite rotation angle of about $\theta=1.74\pi$ (or $\theta=-0.26\pi$). This may come from the slow drift in the trap frequency of our setup between the measurements of $|\psi_-\rangle$ and $|\psi_+\rangle$, which in turn would result in a drift in the rotating speed of the interaction picture and thus cause an effective rotation in the phonon state. As we present in Supplementary Figure 6 of Supplementary Note 2, the supercharges will also be changing under the phonon rotation. The maximizer of the supercharge (strictly speaking, its magnitude) is close to that of the fidelity, but a small discrepancy exists. Here we use the $\theta$ for the highest average state fidelity to calibrate the phonon rotation in $|\psi_+\rangle$, with no such corrections in $|\psi_-\rangle$ since it is measured earlier and the drift in parameters can be smaller. After this correction, the measured supercharges in the main text can be obtained. Also note that once the optimal $\theta$ for $|\psi_+\rangle$ is determined, it is fixed in the data processing. That is, we keep using the same $\theta$ when estimating the error using Monte Carlo sampling as shown in the next section, rather than optimizing a $\theta$ for each simulated result.

\subsection{Error estimation for maximum likelihood method}
\label{app:MC}
We use Monte Carlo sampling to estimate the error in the reconstructed density matrix obtained from maximum likelihood estimation and that in the computed supercharge. These measurement outcomes are computed from the raw experimental data whether the spin is in $|\uparrow\rangle$ or $|\downarrow\rangle$ for each displacement and the red/blue sideband driving at each time point, which is repeated for 400 times (200 times with $\sigma_z$ operations and 200 times without to remove the spin coherence). Note that for each data point, the number of spin-up events follows a binomial distribution whose parameter $p$ can be estimated using the measured frequency $P_{\uparrow}$. Therefore, we can generate new sets of measurement outcomes following these binomial distributions, and then use them to find the most likely density matrix and the corresponding supercharges. We repeat this procedure for 100 times and use the standard deviation of the simulated results to estimate the error of the measurement outcomes.

\subsection{Estimation of preparation and measurement errors for ground state supercharges}
\label{app:error}
The error bars for the measured supercharges using Monte Carlo sampling only consider the uncertainty from finding the best fit for the experimentally prepared ground states. If the state preparation and measurement by themselves are imperfect, clearly there will be a deviation from the ideal values of $\pm 1/\sqrt{2}$. For the adiabatic state preparation, we can numerically integrate the master equation \cite{JOHANSSON20131234} for the linear quench, using a Lindblad term $L[\rho]=1/\tau_d (2a^\dag a \rho a^\dag a - a^\dag a a^\dag a \rho - \rho a^\dag a a^\dag a)$ describing the motional dephasing with an empirical coherence time of about $\tau_d=1.2\,$ms. The supercharges will reduce to $\pm 0.66$ due to this error.

A more severe error source is the slow fluctuation of the trap frequency during the joint spin-phonon state tomography. If such a fluctuation $\Delta\omega_x$ occurs when measuring the blue/red sideband dynamics for the $N=12$ displacement directions, effectively the phonon state will experience a random rotation of $\Delta\omega_x \tau$ where $\tau=200\,\mu$s is the adiabatic evolution time. Such a random rotation among different displacement directions for the same ground state $|\psi_-\rangle$ or $|\psi_+\rangle$ cannot be corrected by a constant rotation angle $\theta$ as done in the previous sections for a fixed drift between the measurements of $|\psi_-\rangle$ and $|\psi_+\rangle$. To estimate its effect, we purposely apply a rotation $\Delta\omega_x \tau$ to the ideal ground state and calculate the supercharges. It turns out that a fluctuation of $2\pi\times 0.5\,$kHz already reduces the supercharges to about $\pm0.52$. In principle, such a slow fluctuation can be suppressed if we calibrate the trap frequency before measuring each data point, which however is too time-consuming for this experiment.

\makeatletter
\renewcommand\@biblabel[1]{#1.}
\makeatother

\bigskip

\textbf{Data Availability:} All the data generated and analysed in this study have been deposited in the Tsinghua cloud database without accession code (\url{https://cloud.tsinghua.edu.cn/f/d9d76ad4c3364b81a9a8/?dl=1}). Contact the corresponding author with any further questions.

\textbf{Code Availability:} The code that support the findings of this study is available from the corresponding authors upon reasonable request.

\textbf{Acknowledgements:} We thank Y.-C. Wang for helpful discussions. This work was supported by the Beijing Academy of Quantum Information Sciences, the Frontier Science Center for Quantum Information of the Ministry of Education of China, and Tsinghua University Initiative Scientific Research Program. Y.-K. W. acknowledges support from National Key Research and Development Program of China (2020YFA0309500) and the start-up fund from Tsinghua University.

\textbf{Competing interests:} The authors declare that there are no competing interests.

\textbf{Author Information:} Correspondence and requests for materials should be addressed to L.M.D.
(lmduan@tsinghua.edu.cn).

\textbf{Author Contributions:} L.M.D. conceived and supervised the project. Y.K.W. did the theoretical calculation. M.L.C., Q.X.M., W.D.Z., Y.J., L.Y., L.H., Z.C.Z. carried out the experiment.  M.L.C., Y.K.W., L.M.D. wrote the manuscript.
\end{document}


\title{Supplementary Information for \\
``Observation of Supersymmetry and its Spontaneous Breaking in a Trapped Ion Quantum Simulator''}

\author{M.-L. Cai}
\thanks{These authors contribute equally to this work}
\affiliation{Center for Quantum Information, Institute for Interdisciplinary Information Sciences, Tsinghua University, Beijing, 100084, PR China}
\affiliation{HYQ Co., Ltd., Beijing, 100176, PR China}

\author{Y.-K. Wu}
\thanks{These authors contribute equally to this work}
\affiliation{Center for Quantum Information, Institute for Interdisciplinary Information Sciences, Tsinghua University, Beijing, 100084, PR China}

\author{Q.-X. Mei}
\affiliation{Center for Quantum Information, Institute for Interdisciplinary Information Sciences, Tsinghua University, Beijing, 100084, PR China}

\author{W.-D. Zhao}
\affiliation{Center for Quantum Information, Institute for Interdisciplinary Information Sciences, Tsinghua University, Beijing, 100084, PR China}

\author{Y. Jiang}
\affiliation{Center for Quantum Information, Institute for Interdisciplinary Information Sciences, Tsinghua University, Beijing, 100084, PR China}

\author{L. Yao}
\affiliation{Center for Quantum Information, Institute for Interdisciplinary Information Sciences, Tsinghua University, Beijing, 100084, PR China}
\affiliation{HYQ Co., Ltd., Beijing, 100176, PR China}

\author{L. He}
\affiliation{Center for Quantum Information, Institute for Interdisciplinary Information Sciences, Tsinghua University, Beijing, 100084, PR China}

\author{Z.-C. Zhou}
\affiliation{Center for Quantum Information, Institute for Interdisciplinary Information Sciences, Tsinghua University, Beijing, 100084, PR China}

\affiliation{Beijing Academy of Quantum Information Sciences, Beijing 100193, PR China}

\author{L.-M. Duan}
\email{lmduan@mail.tsinghua.edu.cn}
\affiliation{Center for Quantum Information, Institute for Interdisciplinary Information Sciences, Tsinghua University, Beijing, 100084, PR China}

\maketitle

\makeatletter
\renewcommand{\figurename}{Supplementary Figure}
\makeatother

\section*{Supplementary Note 1: Supersymmetry in Quantum Rabi model}
A quantum mechanical system is supersymmetric if we can define some Hermitian supercharges $Q_1$, $Q_2$, $\cdots$, $Q_N$ such that $\{Q_i,\, Q_j\} = 2 H \delta_{ij}$,
where $H$ is the Hamiltonian of the system \cite{SUSYQM2004}. By definition, all the supercharges commute with $H=Q_i^2$, so that $Q_i$'s are conserved quantities.

The simplest case is the $N=2$ SUSY QM with two supercharges. Here we can define a Witten parity operator $K$ satisfying $K^2=I$ and $\{K,\, Q_i\} = 0$ ($i=1,\,2$).
It can be shown that for the $N=2$ SUSY QM we can choose $Q_2=\pm i K Q_1$, so that we only need to consider one supercharge \cite{SUSYQM2004}. From the above definitions, we can see that $[H,\, K] = 0$, hence $K$ is also conserved. We can thus split the whole Hilbert space into $\mathcal{H}=\mathcal{H}_+\oplus\mathcal{H}_-$ where $\mathcal{H}_\pm$ is the eigenspace of $K$ with eigenvalue $\pm 1$ which corresponds to the bosonic and the fermionic states, respectively.

Now we consider the QRM model [Eq.~(1) of the main text], which is supersymmetric at two sets of parameters \cite{Hirokawa2015}: $g=0$ and $\omega_s=\omega$; or $\omega_s=0$.

\subsection{$g=0$ and $\omega_s=\omega$}
In this case, the Hamiltonian $H=\omega\sigma_z/2+\omega(a^\dag a + 1/2)$ is just a spin and a bosonic mode without any interaction. It is easy to check that $Q=\sqrt{\omega}(a\sigma_++a^\dag \sigma_-)$ and $K=\sigma_z$ satisfy all the above definitions. Here we can see that the spin-up and the spin-down states with different phonon numbers give the whole spectrum of the bosonic and the fermionic states, and the mapping from spin-up/spin-down to bosonic/fermionic states is arbitrary: We can simply define $K\to -K$ to reverse the Witten parity operator.

Let us choose $|\downarrow\rangle|n\rangle$ as the bosonic states and $|\uparrow\rangle|n\rangle$ as the fermionic states. Here we have a unique ground state $|\downarrow\rangle|0\rangle$ with energy $E_0=0$, and the higher levels $|\downarrow\rangle|n+1\rangle$ and $|\uparrow\rangle|n\rangle$ are degenerate with energy $E_{n+1}=(n+1)\hbar\omega$ ($n=0,\,1,\,\cdots$). The supercharge $Q$ transforms the degenerate bosonic and fermionic states into each other with a nonzero normalization factor, and it annihilates the unique ground state.

\subsection{$\omega_s=0$}
In this case the Hamiltonian is given by
\begin{equation}
H = \omega \left(a^\dag a + \frac{1}{2} \right) + g \sigma_x (a + a^\dag) + \frac{g^2}{\omega}.
\end{equation}
To demonstrate the SUSY structure more clearly, we follow Ref.~\cite{Hirokawa2015} to perform a unitary transform
\begin{equation}
U_g = \frac{1}{\sqrt{2}} \left(
\begin{array}{cc}
V_- & -V_+\\
V_- & V_+
\end{array}
\right),
\end{equation}
where $V_\pm = \exp[\pm g/\omega(a^\dag - a)] = D(\pm g/\omega)$ is the displacement operator of $\pm g/\omega$. It is straightforward to verify that
\begin{equation}
U_g^\dag H U_g = \omega I\otimes \left(a^\dag a + \frac{1}{2}\right).
\end{equation}

A supercharge in this transformed frame can be given by
\begin{equation}
U_g^\dag Q U_g = \sigma_x \otimes \sqrt{\omega(a^\dag a + 1/2)},
\end{equation}
with the Witten parity operator $U_g^\dag K U_g = \sigma_z \otimes I$.
Moving back to the original frame, we get
\begin{align}
Q =& U_g \sigma_x \otimes \sqrt{\omega(a^\dag a + 1/2)} U_g^\dag \nonumber\\
=& -\frac{\sigma_z}{2}\otimes \left[V_+ \sqrt{\omega(a^\dag a + 1/2)} V_+ + V_- \sqrt{\omega(a^\dag a + 1/2)} V_-\right]  \nonumber\\
& - i\frac{\sigma_y}{2} \otimes \left[V_+ \sqrt{\omega(a^\dag a + 1/2)} V_+ - V_- \sqrt{\omega(a^\dag a + 1/2)} V_- \right],  \label{eq:q1}
\end{align}
and $K=\sigma_x$. In the main text, we have taken out the factor $\sqrt{\omega}$ from the definition of $Q$ to make it dimensionless. Then we have $H=\omega Q^2$.

The ground state in the transformed frame has energy $E_0=\omega/2$ when the phonon number is zero. If we further require the states to be the eigenstates of $Q$, we see that, in the transformed frame, the two ground states can be chosen as $|\pm\rangle|0\rangle$, with eigenvalues of $\pm 1/\sqrt{2}$ for the supercharge. Now if we move back to the original frame, the two ground states can be expressed as $U_g |\pm\rangle|0\rangle = (|+\rangle|-g/\omega\rangle \mp |-\rangle|g/\omega\rangle) / \sqrt{2}$.

\section*{Supplementary Note 2: Complete experimental data for partial spin-phonon state tomography}
\begin{figure}[htbp]
   \includegraphics[width=\linewidth]{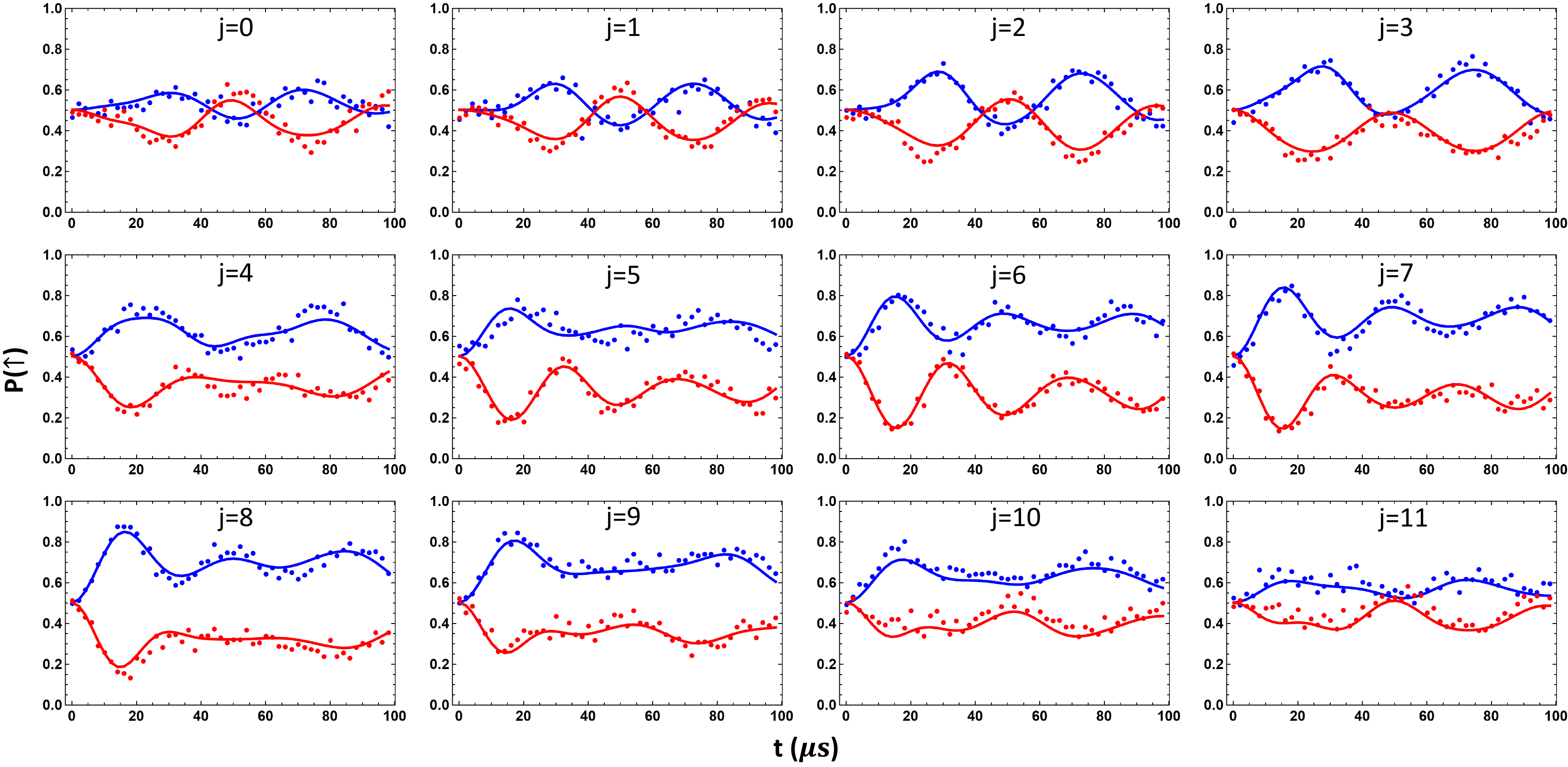}
   \caption{Experimental data (dots) and the theoretical prediction based on the best-fitted density matrix (solid curves) when driving the blue or red sidebands for the $|\psi_-\rangle$ state projected to $\sigma_y=\pm 1$. We choose $\omega=2\pi\times 10\,$kHz and $g_m=2\pi\times 5.43\,$kHz. The displacement operators $D(\beta_j)$ are characterized by $\beta_j=i\beta e^{2\pi i j / N}$ where $\beta=0.687$, $N=12$ and $j=0,\,1,\,\cdots,\, N-1$.}
   \label{fig:down_y}
\end{figure}

\begin{figure}[htbp]
   \includegraphics[width=\linewidth]{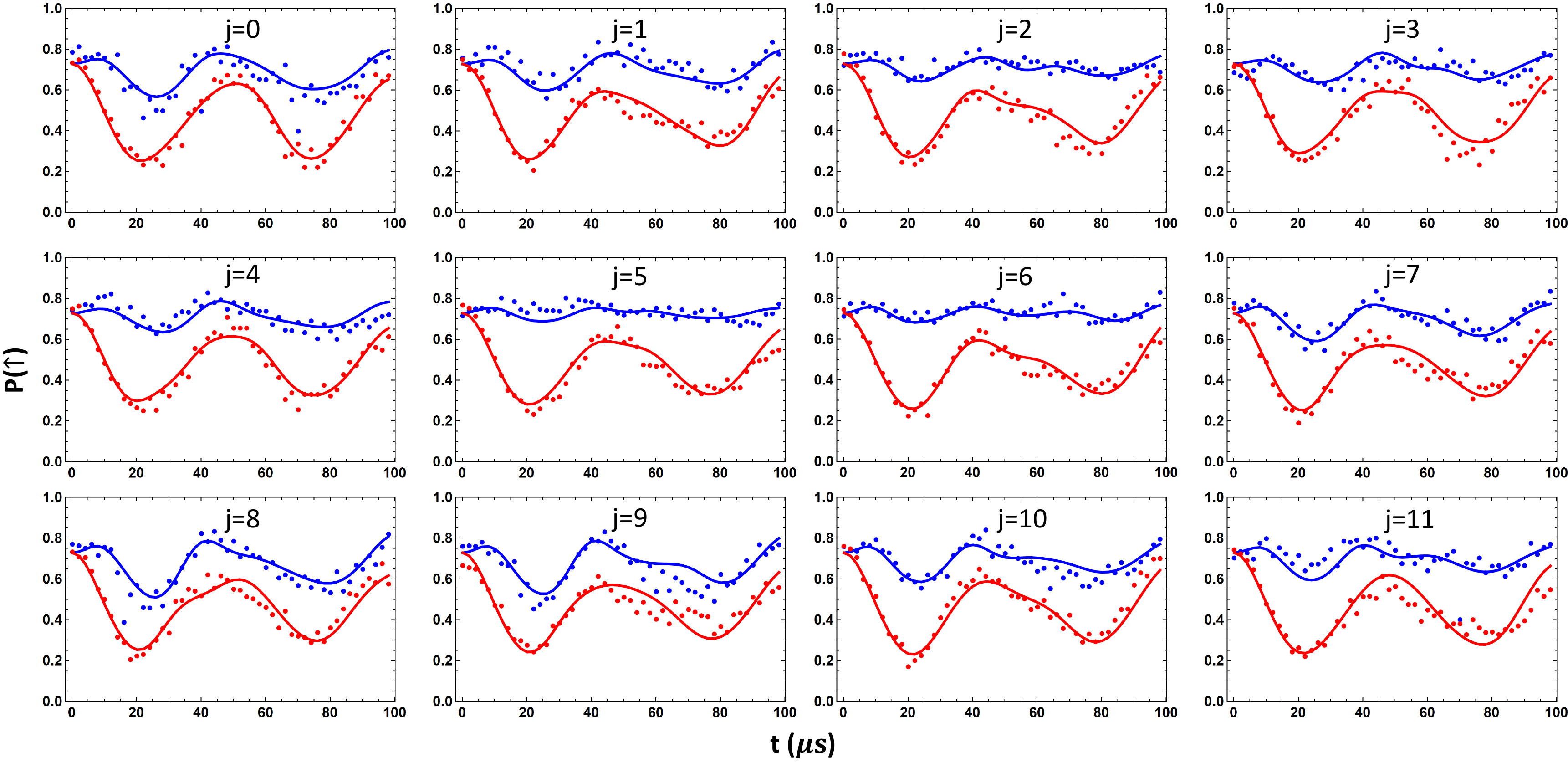}
   \caption{Experimental data (dots) and the theoretical prediction based on the best-fitted density matrix (solid curves) when driving the blue or red sidebands for the $|\psi_+\rangle$ state projected to $\sigma_z=\pm 1$. We choose $\omega=2\pi\times 10\,$kHz and $g_m=2\pi\times 5.43\,$kHz. The displacement operators $D(\beta_j)$ are characterized by $\beta_j=i\beta e^{2\pi i j / N}$ where $\beta=0.687$, $N=12$ and $j=0,\,1,\,\cdots,\, N-1$.}
   \label{fig:up_z}
\end{figure}

\begin{figure}[htbp]
   \includegraphics[width=\linewidth]{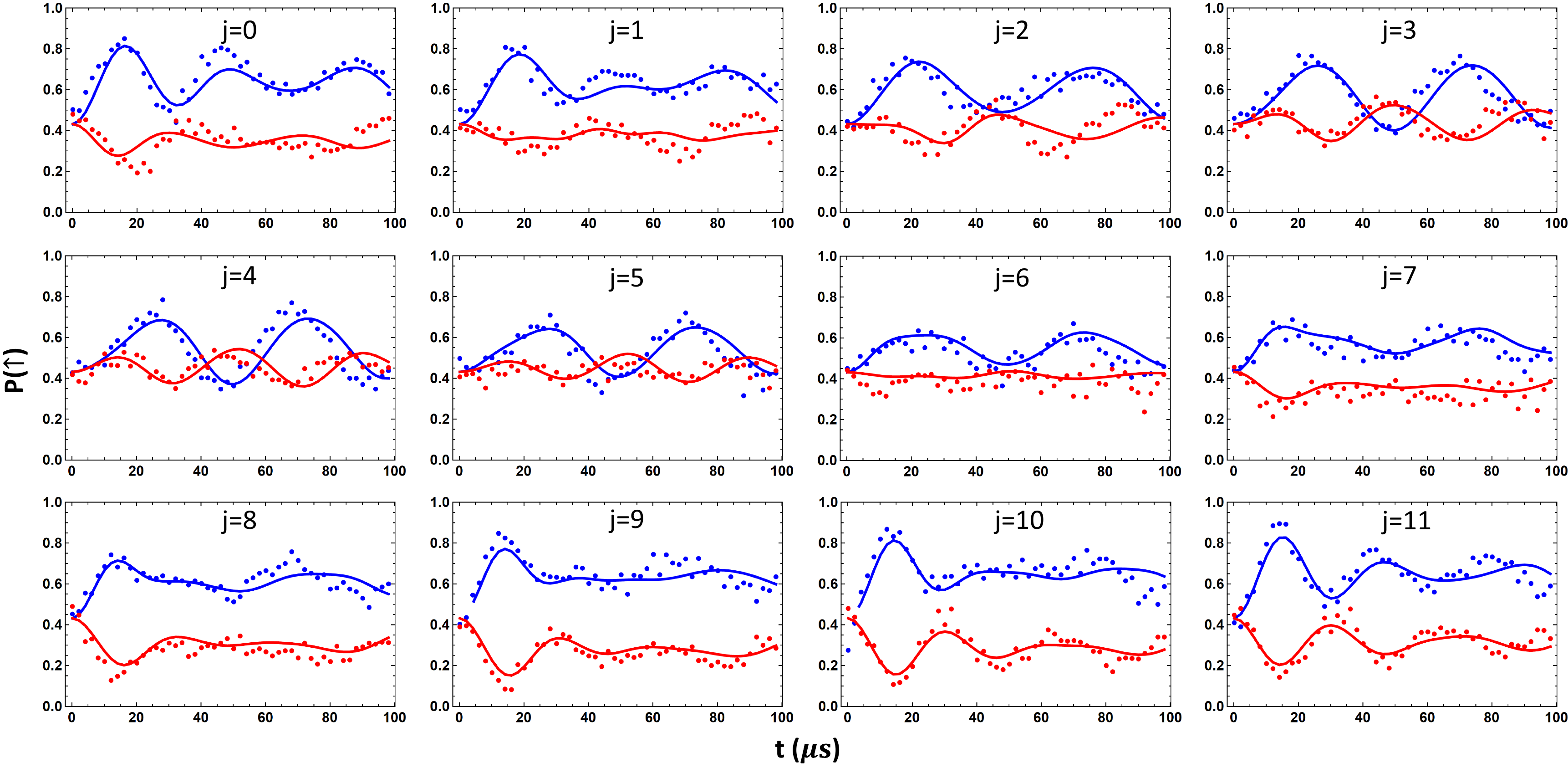}
   \caption{Experimental data (dots) and the theoretical prediction based on the best-fitted density matrix (solid curves) when driving the blue or red sidebands for the $|\psi_+\rangle$ state projected to $\sigma_y=\pm 1$. We choose $\omega=2\pi\times 10\,$kHz and $g_m=2\pi\times 5.43\,$kHz. The displacement operators $D(\beta_j)$ are characterized by $\beta_j=i\beta e^{2\pi i j / N}$ where $\beta=0.687$, $N=12$ and $j=0,\,1,\,\cdots,\, N-1$.}
   \label{fig:up_y}
\end{figure}

In Supplementary Figure~\ref{fig:down_z}, Supplementary Figure~\ref{fig:down_y}, Supplementary Figure~\ref{fig:up_z} and Supplementary Figure~\ref{fig:up_y} we present the complete experimental data for measuring the joint spin-phonon state by projecting $|\psi_\pm\rangle$ onto $\sigma_z=\pm 1$ or $\sigma_y=\pm 1$, as well as the theoretical results for the best fitted joint density matrices.

\begin{figure}[htbp]
   \includegraphics[width=\linewidth]{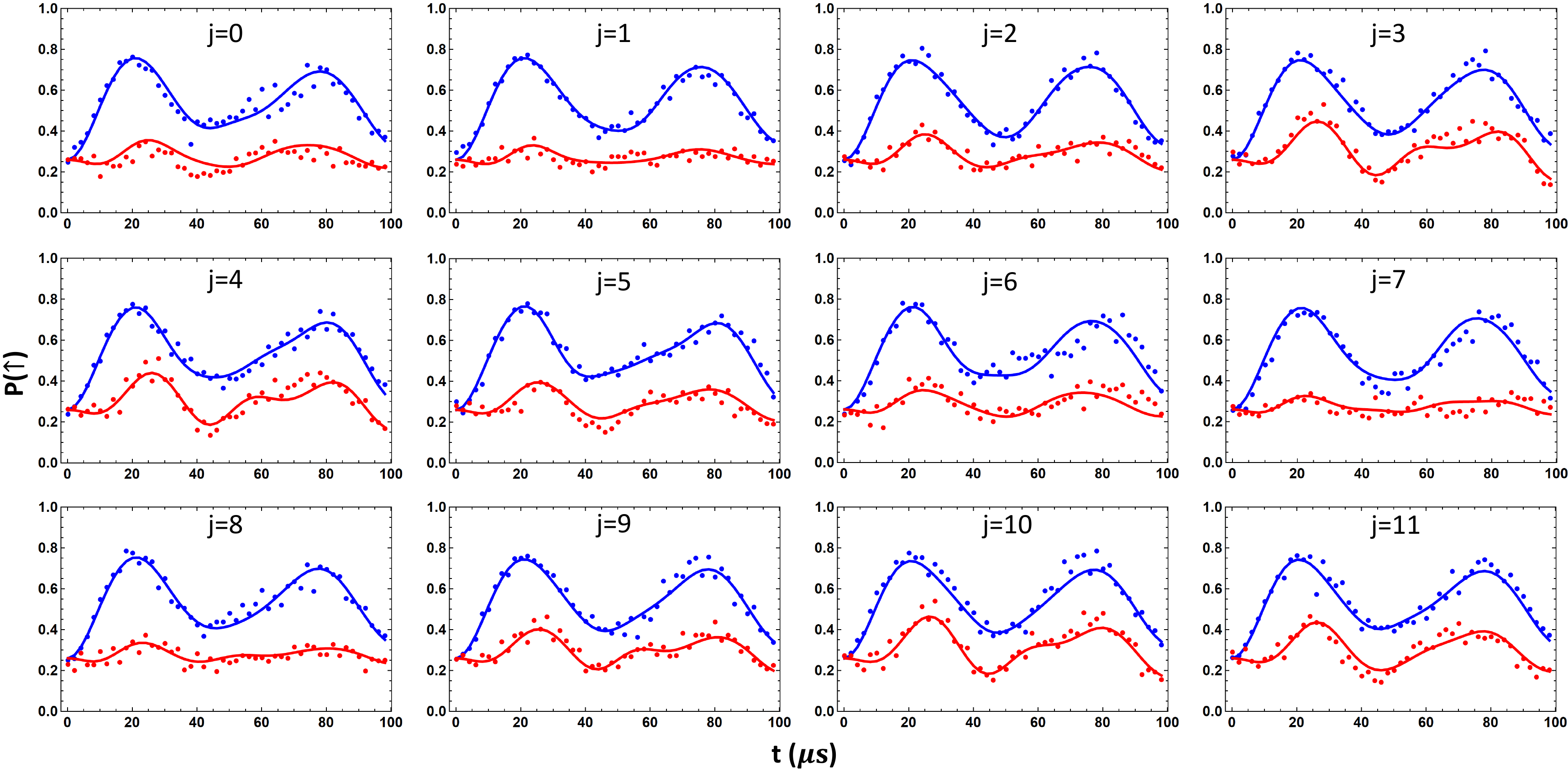}
   \caption{Experimental data (dots) and the theoretical prediction based on the best-fitted density matrix (solid curves) when driving the blue or red sidebands for the $|\psi_-\rangle$ state projected to $\sigma_z=\pm 1$. We choose $\omega=2\pi\times 10\,$kHz and $g_m=2\pi\times 5.43\,$kHz. The displacement operators $D(\beta_j)$ are characterized by $\beta_j=i\beta e^{2\pi i j / N}$ where $\beta=0.687$, $N=12$ and $j=0,\,1,\,\cdots,\, N-1$.}
   \label{fig:down_z}
\end{figure}

\begin{figure*}[htbp]
   \includegraphics[width=0.8\linewidth]{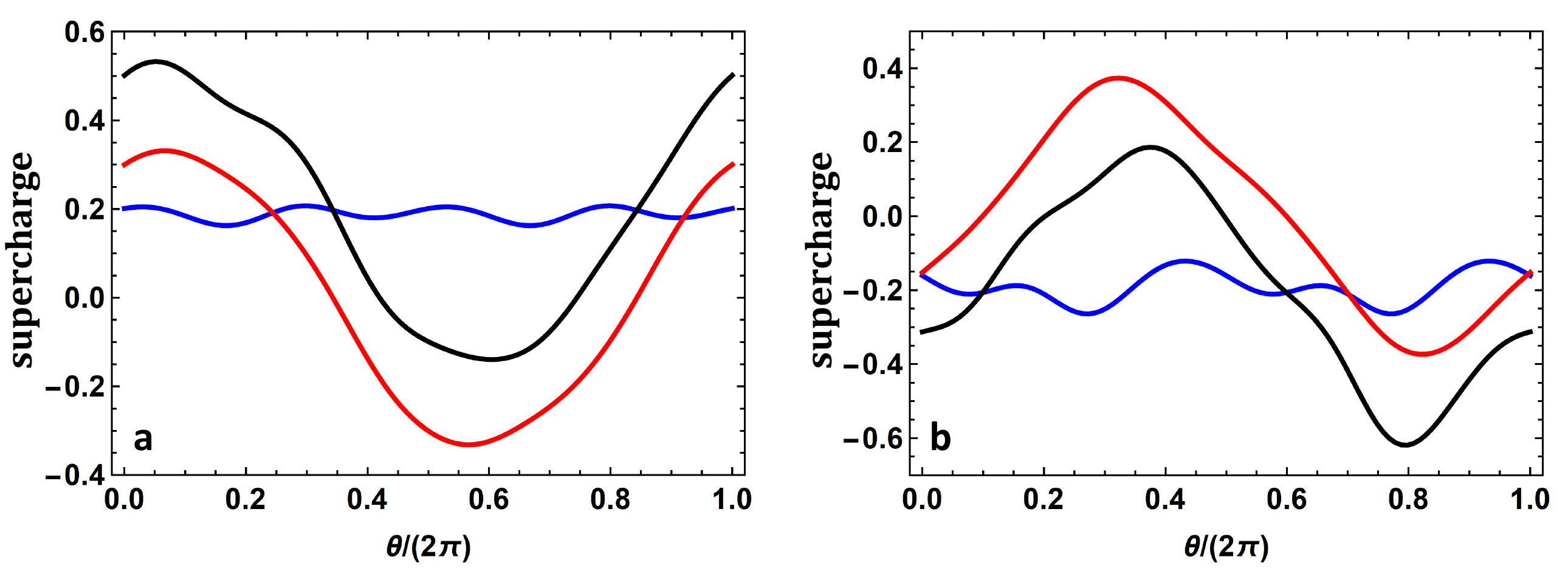}
   \caption{Expectation values for $\sigma_z\otimes A$ (blue), $\sigma_y\otimes B$ (red) and the total supercharge (black) for \textbf{a} $|\psi_-\rangle$ and \textbf{b} $|\psi_+\rangle$.}
   \label{fig:supercharge_rotate}
\end{figure*}

In Supplementary Figure~\ref{fig:fid_rotate} and Supplementary Figure~\ref{fig:supercharge_rotate} we present the experimental data used to calibrate a phonon state rotation angle between the measurement of $|\psi_+\rangle$ and $|\psi_-\rangle$ as described in Methods.

\begin{figure*}[htbp]
   \includegraphics[width=0.8\linewidth]{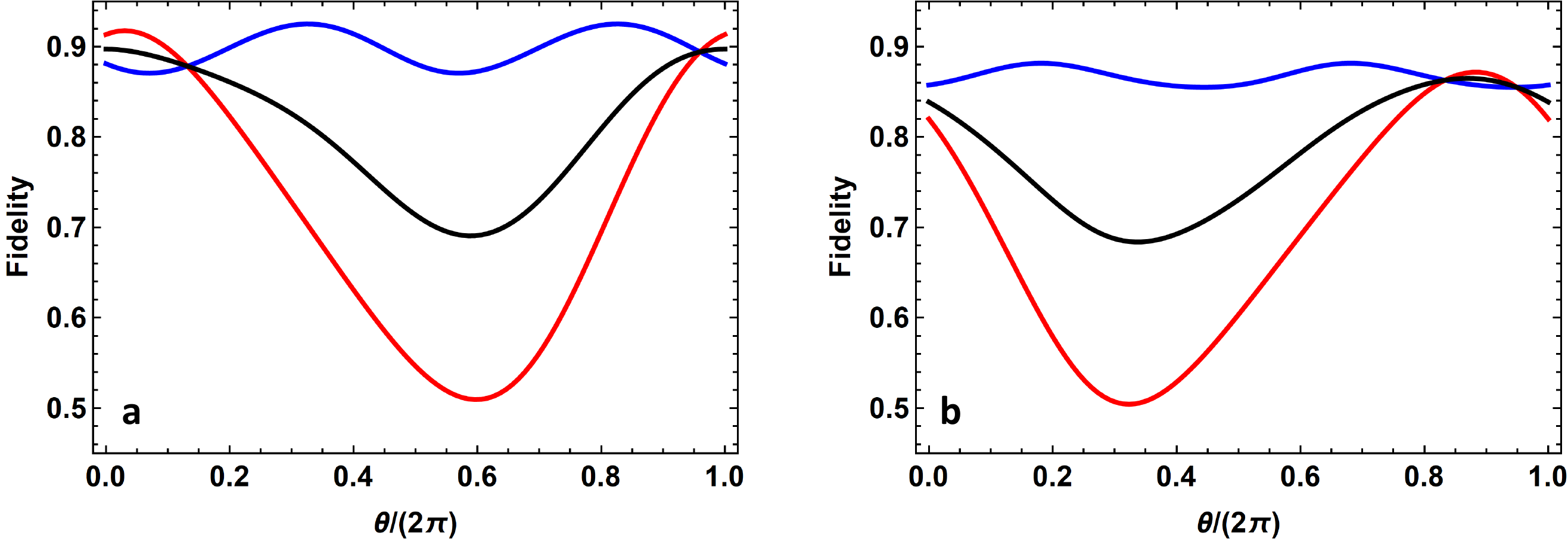}
   \caption{Fidelity for the spin-phonon state projected to $\sigma_z=\pm 1$ (blue), $\sigma_y=\pm 1$ (red) and their average (black) for \textbf{a} $|\psi_-\rangle$ and \textbf{b} $|\psi_+\rangle$.}
   \label{fig:fid_rotate}
\end{figure*}
